# Bypassing the single junction limit with advanced photovoltaic architectures

Larry Lüer[1], Marius Peters[2], Dan Bornstein[1], Vincent M. Le Corre[1], Karen Forberich[2], Dirk Guldi[3], Christoph J. Brabec[1,2]

[1] Institute of Materials for Electronics and Energy Technology (i-MEET), Friedrich-Alexander-Universität Erlangen-Nürnberg, Martensstrasse 7, 91058 Erlangen, Germany

[2] High Throughput Methods in Photovoltaics, Forschungszentrum Jülich GmbH, Helmholtz Institute Erlangen-Nürnberg for Renewable Energy (HI ERN), Immerwahrstraße 2, 91058 Erlangen, Germany

[3] Department of Chemistry and Pharmacy, Egerlandstr. 3, 91058 Erlangen, Germany

## Abstract

In single-junction photovoltaic (PV) devices, the maximum achievable power conversion efficiency (PCE) is mainly limited by thermalization and transmission losses, because polychromatic solar irradiation cannot be matched to a single bandgap. Several concepts are being investigated to reduce these losses, such as the classical vertical multijunction cells, "lateral "tandem cells, and multi-exciton generation in the form of photon up- and down-conversion. While in theory, efficiencies exceeding 90% are possible (Landsberg or thermodynamic limit), there are severe practical limitations in terms of processability, cost, and spectral sensitivity. Here, we present a simulation environment based on Bayesian Optimization that is able to predict and optimize the electrical performance of multi-junction architectures, both vertical and lateral, in combination with multi-exciton materials. With respect to vertical stacks, we show that by optimizing bandgap energies of multi-exciton generation (MEG) layers, double junctions can reach efficiencies beyond those of five-junction tandem devices (57%). Moreover, such combinations of MEG and double junction devices would be highly resilient against spectral changes of the incoming sunlight. We point out three main challenges for PV material science to realize such devices. With respect to lateral architectures, we show that MEG layers might allow reducing nonradiative voltage losses following the Energy Gap Law. Finally, we show that the simulation environment is able to use machine learned quantitative structure-property relationships obtained from high-throughput experiments to virtually optimise the active layer (such as, the film thickness and the donor-acceptor ratio) for a given architecture. The simulation environment thus represents an important building block towards a digital twin of PV materials.

## 1    Introduction

Photovoltaic (PV) technology is considered one of the pillars of the transition towards a sustainable society.[1] Power conversion efficiency (PCE) values of PV systems currently on the market range from 23.7% (CIGS) to 26.8% (c-Si), while emerging PV technologies such as organic photovoltaics and perovskites reach 18.2% and 25.7%, respectively.[2] The reason for these limited efficiencies is the restriction to a single semiconductor bandgap; photons above the bandgap are absorbed but lose excess energy (thermalization losses), while photons below the bandgap are not absorbed by the semiconductor (transmission losses). The Single Junction Limit in an ideal semiconductor material with abrupt absorption onset, presenting only unavoidable radiative losses, is close to 33%.

Various strategies are being pursued to surpass the Single Junction Limit in PV technology. Multijunction architectures combine several semiconductors with various bandgaps such that thermalization and transmission losses can be greatly reduced (see Figure 1a, right). In fact, the current world record cell, according to the chart of the U.S. National Renewable Energy Labs (NREL), is a multi-junction cell reaching 47.6% under 665 suns.[2] However, the fabrication of multiple junctions is technologically demanding, such that multijunction cells are too expensive to compete with existing technologies except in outer space where power densities are extremely important. Moreover, the single sub-cells in multi-junction architectures are usually connected in series (see Figure 1b, right); Kirchhoff's rules then require current matching between all sub-cells of the multi-junction, which must be carefully optimized for a given light spectrum but severely suffers if the spectrum changes, for example, towards the evening, depending on albedo and latitude of deployment.

Spectral sharing strategies are also studied, by applying optical elements such as beam splitters[3] or phase masks, either planar[4] or focusing,[5] which lead to a lateral splitting of the solar spectrum. Thus, "rainbow solar cells"[6] can be constructed by a lateral architecture of the multijunction. These rainbow cells can be built more easily than vertical architectures, however, the Kirchhoff rules must still be respected.

An alternative approach to multijunction cells is the use of materials that present multi-exciton generation (MEG) phenomena, the most prominent ones being photon up-conversion or down-conversion (UC and DC, respectively). UC materials that convert (at least) two low-energy photons into one high-energy photon present long lifetimes of mobile excitons such that annihilation processes can proceed at high yields, creating excited states of double energy which can be harvested by sensitizers.[7] UC materials typically require concentrated light to outperform first order deactivation processes. Assuming typical kinetic parameters, up-conversion efficiencies are predicted to be close to 20-30 % under 1 sun[7] and can be boosted significantly even at moderate concentration factors; this shows the potential to go beyond the current limit of 7%.[8]

DC materials, on the other hand, form primary excited states that can split into two lower excited states. The most prominent process is singlet fission, by which one singlet state converts into two triplet states that can stabilize in different spatial regions of the conjugated system.[9] The triplet states can be converted into charge carriers via exciton dissociation at the interface between the DC and semiconductor (SC) phases (CT coupling, see Figure 1a, middle). However, level matching with common semiconductor materials has proven difficult as there are constraints on the chemical structure of DC materials leading to the required low-lying triplet states.[10] Even when levels would favour a charge transfer across the DC-SC interface, typical diffusion lengths of triplet excitons do not match typical light penetration depths, a situation known in donor:acceptor blends in organic PV. In the latter field, the issue has been addressed by forming interpenetrating networks ("bulk heterojunctions") between donor and acceptor with domain sizes below the exciton diffusion lengths, which allows layer thicknesses on the order of the light penetration depths.[11] This approach has been tried also for DC:SC systems; however, it was found that the primary excitations (singlets) can transfer from the DC into the SC phase before the fission process takes place thus re-introducing the undesired thermalization; moreover, a quenching of triplet states with charged states was observed. For these reasons, the alternative route of radiative transfer of triplet energy has been elaborated, see Figure 1a, left. By exploiting spin-orbit coupling in PbS nanoparticles, up to 90% of triplet-photon conversion yields have been reached.[12]

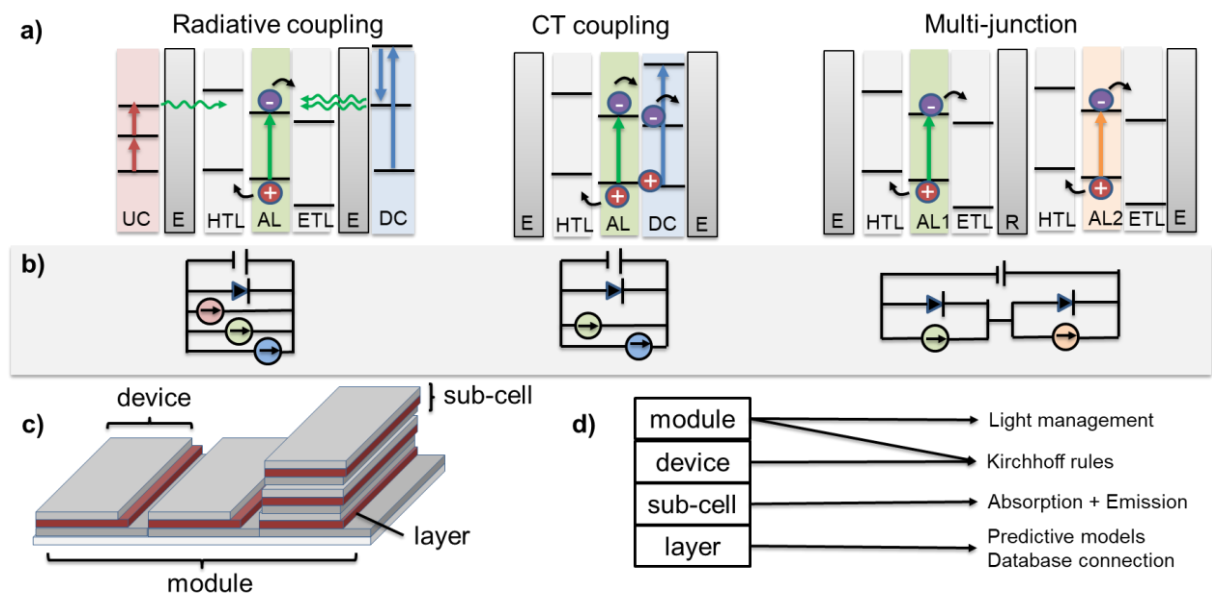

Figure 1. a) Architectures, level schemes and transfer processes for radiatively coupled and CT coupled up-conversion and down-conversion materials (UC and DC, respectively), and a double junction system. Abbreviations for the layers are: HTL: hole transporting layer, ETL: electron transporting layer, AL, active layer, E: semi-transparent electrode, R: recombination layer. B) corresponding equivalent circuits, c) Unique schematic representation of both vertical and lateral multi-junction stacks, d) Functionalities of the corresponding Python classes in the BOAR simulation and optimization environment. For more details, see ESI Figure S1.

In an equivalent circuit picture (see Figure 1b), MEG processes are connected in parallel with the SC of the corresponding single junction, meaning that the photons generated in the UC and DC layers are simply added to the photons absorbed in the SC. This has the decisive advantage that according to Kirchhoff's rules, current matching is not required.[18] MEG-based PV systems should therefore be more resilient against spectral changes of the sunlight over the day & year and along the lifeline (soiling; colouring of packaging material) than multi-junction systems. The disadvantage of MEG systems, from a theoretical perspective, is that the energy ratio of DC and UC against SC must be about a factor of two, dictated by quantum conservation, while multi-tandem approaches allow in principle arbitrarily close energy levels. As ~80% of the photons of the solar spectrum are contained within the two octaves from 400 to 1600 nm, no more than three bandgaps (including UC,DC, and SC layers) can be accommodated using MEG principles. Therefore, multi-junction systems can theoretically reach higher PCE values, however at the price of low resilience against spectral changes.[18]

Organic semiconductors represent a very versatile class of materials, covering a large range of level energies and optical bandgaps. In contrast to inorganic, band-like materials, organic semiconductors form local excited states that cause selective absorption in energetically limited absorption bands, alternating with transmissive regions along the energy scale. This feature can be interesting for spectral sharing concepts for green houses[13] or building integration[14] but also for novel architectures combining vertical and lateral multi-junction systems with MEG layers, thus creating highly efficient but still spectrally resilient PV systems.

The multitude of possible architectural choices, together with the design freedom of organic semiconductors, creates a large design space in which experimentation will take place. It is therefore mandatory to exploit prior knowledge to reduce the design space as much as possible, in order to direct experimentalists towards the most promising combination of materials for a given architectural choice. This calls for an integrated simulation environment able to perform global optimization tasks within a variety of architectural choices (lateral versus vertical stacks combined with MEG layers inside or outside the sub-cells, according to Figure 1a) and photophysical transfer mechanisms (radiative, resonant, charge transfer), exploiting available knowledge about the deployed materials. Here, we present the simulation environment BOAR (Bayesian Optimization for Automatic Research) able to address these tasks. BOAR is a python library designed for any kind of optimization problem or fitting problem with op to ten parameters. In this study, we have used it to predict the electrical performance of lateral or vertical multi-stack architectures by using quantitative structure property relationships obtained by high throughput experimentation.[15]

After describing the workflow and operation modes of the simulation environment, we present results for ideal cases, which is useful to know the maximum achievable performance in terms of efficiency and spectral resilience. Finally, we use the simulation environment to predict the performance of multi-stack architectures from the real performance of single-stack systems.

## 2   Methods

Figure 1c describes the ontology that the BOAR simulation and optimization environment uses to treat lateral and vertical architectures on a common footing. On the module level, light management by phase masks is simulated along the direction of the lateral stacks. To model the effect of the phase masks, we assume a series of long pass filters (sigmoidal dashed lines in ESI, Figure S3) that in combination act as adjustable bandpass filters for the solar irradiation onto each of the single devices that make up the module (AM1.5 – AM5.0, see blue and orange spectra in Figure 1d, respectively). This amounts to assuming perfect spectral selectivity except for the transition region between two adjacent spectral bands, a condition which is not perfectly fulfilled for uncollimated light.[4] On the device and module levels, a series connection is assumed and Kirchhoff's rules are enforced. Devices can consist of an arbitrary number of sub-cells forming a multi-junction, where each sub-cell is a complete photovoltaic cell. According to Figure 1a, MEG layers can either be coupled radiatively or via charge transfer (CT) coupling. Across all sub-cells of a device (vertical stack), light absorption, reflection, and emission are calculated assuming geometrical optics, disregarding interference. Finally, on the layer level, the electrical performance parameters, namely open circuit voltage, short circuit current, and fill factor ($V_{OC}$, $J_{SC}$, FF, respectively) are calculated. Three modes are available (see ESI Fig. S1): in mode "boxcar", unity absorptance is assumed between the lower and higher optical band limits, $E_{opt}^0$ and $E_{opt}^1$, respectively, and zero otherwise. In mode "spectral", a general spectral model for carbon-based semiconductors is assumed with adjustable electron-phonon coupling, ratio of amorphous to ordered phases and suppression or enhancement of the vibronic origin by H or J-type aggregation (Spano model[16]). Spectral models for common electron donor and acceptor materials are available. In the "spectral" mode, $E_{opt}^0 = c_1^{LBG} - w_1^{LBG}$, where $c_1^{LBG}$ and $w_1^{LBG}$ are the energy and bandwidth, respectively, of the exciton band of the low energy component in the donor:acceptor system. This method approximates the generally accepted method to obtain the optical bandgap from the low energy inflection point of the external quantum efficiency.[17] From the absorptance, $J_{SC}$ is calculated by multiplying the excitation current with the internal quantum efficiency (IQE). In both "boxcar" and "spectral" modes, $V_{OC}$ is obtained by $V_{OC} = E^0_{opt} - \Delta V_{OC}$, where $\Delta V_{OC}$ is the voltage loss, considering both radiative and non-radiative contributions, and the power conversion efficiency (PCE) is obtained by multiplying $J_{SC}$ and $V_{OC}$ with FF and dividing by the total incoming power.

In mode "predictive", we use trained Gaussian Process Regressors (GPR) from high throughput experiments to predict $J_{SC}$, $V_{OC}$, and FF from spectral shapes of the UV-Vis spectra of the active layer. For details, we refer to ESI, part A. As these GPR models were trained on devices with reflective bottom electrodes, light interference effects are implicitly considered. Using GPR as surrogate models allows us to perform virtual optimizations of layer parameters such as thickness or D:A ratios for a given vertical or lateral architecture. Without GPR surrogates, thickness-dependent simulations would involve drift-diffusion simulations, and simulations of the effect of D:A ratios, due to the disordered nature of the active layers, would even involve kinetic Monte Carlo simulations which are not suitable to be coupled to a Bayesian Optimization scheme due to the computation time of each simulation. Thus, only the GPR surrogates allow rapid predictions which can be used by the Bayesian Optimizer to encounter the global optimum within the allowed search space. On an Intel i7 processor of 12$^{th}$ generation, optimization of 6 parameters in the "predictive" mode takes less than 2 minutes, which includes calculation of the posterior distribution for uncertainty quantification.

## 3    Results

### 3a    Ideal systems

By analysing ideal systems, we encounter the theoretical limits that the chosen architectures can exhibit. We start with boxcar absorbers having a fixed FF of 0.8 and voltage losses of $\Delta V_{OC} = 0.2\ V$, typical for inorganic systems. In Figure 2a, we show PCE as function of the air mass (related to the azimuthal angle of the sun against the normal) for multi-junction architectures. Varying the air mass quantitatively describes spectral changes from midday to evening, however, it is also a qualitative measure for spectral changes due to other reasons (soiling, encapsulant coloring, latitude, see ESI, Fig. S2). In order to assess the resilience of the PV systems against spectral changes, we have used BOAR to optimize the available bandgap energies at AM1.5 and then simulated PCE as function of the air mass (from AM1.5-5.0) without re-optimizing the bandgap energies. For a single junction (blue line and symbols), the well-known limit of approximately 33% is recovered at AM1.5. We observe a slight rise towards 34% under AM5.0 which is due to the spectral filtering effect of the larger air mass predominantly at short wavelengths, narrowing the spectral shape below that of a blackbody emitter, thus making the solar spectrum slightly more "monochromatic" which leads to smaller thermalization losses in the PV device. Multijunction vertical devices are shown in orange, green, red, and violet for 2, 3, 4, and 5 junctions, respectively. At AM1.5, PCE values of up to 55% can be reached for the 5-junction system. However, we pay the price of low resilience against spectral changes: due to Kirchhoff's rules, reducing the excitation current in the top device (responsible for harvesting the short wavelengths, suppressed under AM 5.0) means reducing the current of all other sub-cells too. At AM5.0, the requirement of current matching imposed by the serial connection depresses the PCE for a 5-junction device to 39%, below that of a 2-junction device.

Figure 2 b shows the PCE trends for the corresponding lateral architectures. For lateral light management, we assumed bandpass filters with an edge selectivity of 10 nm (see ESI, Fig. S2).[4] Under this assumption, PCE values and spectral sensitivities follow the same trend as in the vertical architectures. However, the maximum PCE values are slightly lower in the lateral than in the vertical architectures, owing to the assumed imperfect spectral selectivity. The effect becomes stronger if more lateral junctions are present, compare curves of same colour in Figure 2a and b, respectively.

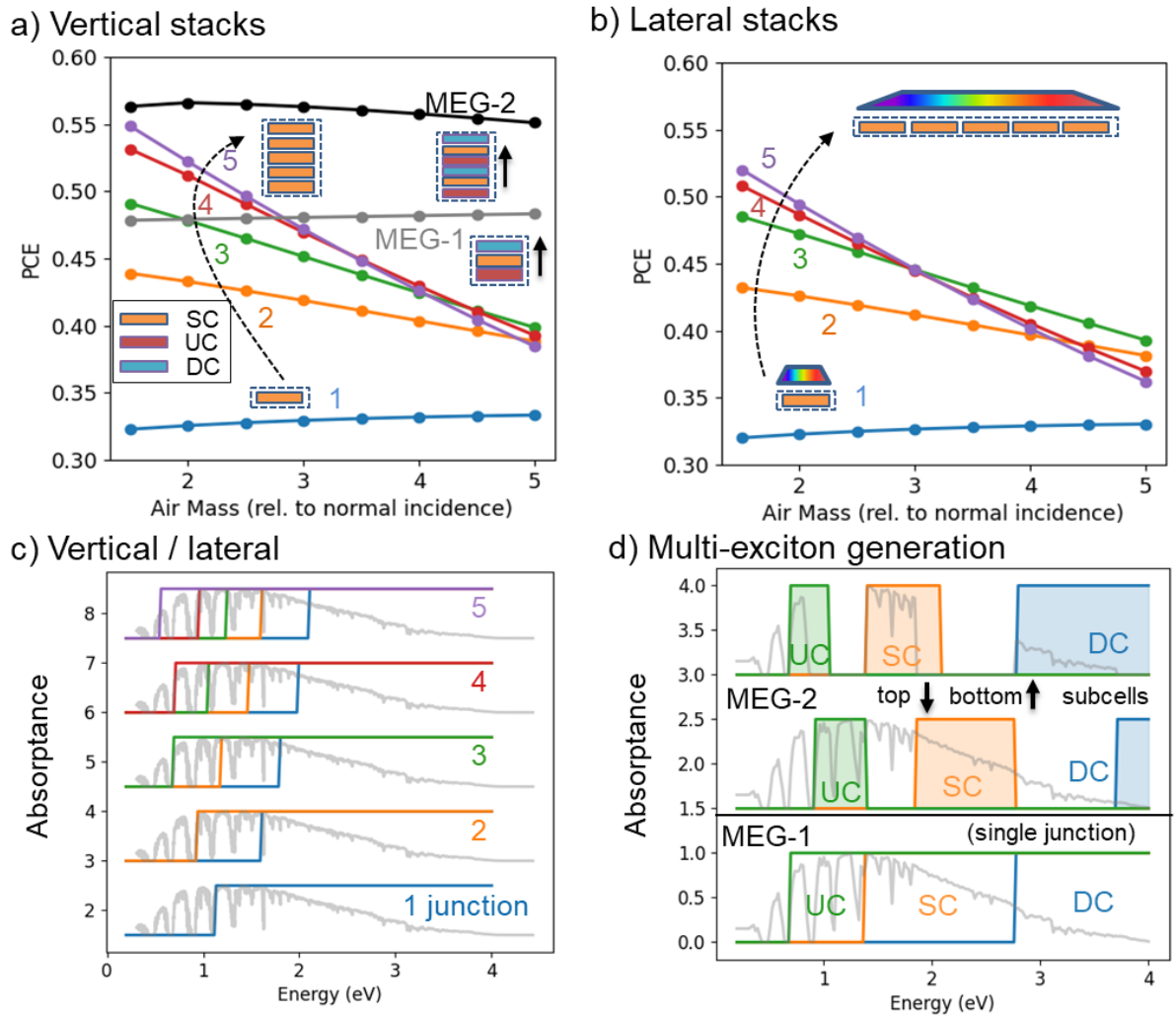

*Figure 2. Ideal material systems. a) PCE against air mass (AM) for vertically stacked multi-junction systems, with the number of junctions given in the inset, and for single junction and double junction MEG systems (gray and black lines, respectively). All systems were optimized at AM1.5. b) Same as a) for laterally aligned multi-junctions. A spectral selectivity of 10 nm has been assumed for the spectral filters as defined in Figure 2 c) Optimal bandgaps of ideal systems for both vertical and lateral alignments. Abrupt absorption onset and total absorption are assumed for all semiconductors. The normalized AM1.5 spectrum, given in units of $W/eV/m^2$ is shown as gray lines. Data from each individual junction are shifted by 1.5 for clarity of presentation. d) Same as c for MEG systems. MEG-1 is a single-junction stack comprising both DC and UC materials. For the double junction system MEG-2, staggered half octave boxcar absorbers have been assumed. For the bottom layer, the incident residual AM1.5 spectrum after transmission through the top cell is shown as gray line.*

As shown in Figure 1b, MEG systems are electrically parallelly connected to the photovoltaic cells irrespective whether in CT or radiative mode; therefore, even a complete loss of one channel would not block overall current generation. The gray line in Figure 2a shows the evolution of an ideal single junction system comprising both UC and DC, charge coupled to the SC layer. As shown in Figure 1a, the bandgaps of both UC and DC must be one octave (factor of 2)[18] away from the bandgap of the SC layer which is thus the only free parameter that can be optimized. Due to this constraint, the ideal single junction MEG system stays slightly below the triple junction stack at about 48% (compare gray and green curves at AM1.5, respectively), which also has three bandgaps, but which can be optimized individually without needing to conserve the octave offset. Most importantly, for AM3 and beyond, the single junction MEG system outperforms even the 5-junction vertical stack, showing high resilience against spectral changes.

For the single junction MEG system (MEG-1 shown in Fig. 3d), we have assumed abrupt boxcar absorption from the bandgap up to infinite energy, a good approximation for inorganic semiconductors. In contrast, organic semiconductors, due to the localized nature (Frenkel excitons) of their excited states, exhibit selective absorption bands alternating with transmissive regions. This opens the opportunity to construct a hypothetical two-junction device in which each sub-cell contains UC, DC, and SC layers. As shown in Figure 2d, to provide illumination to the bottom sub-cell while still respecting the octave offset requirements for both sub-cells, means that all UC, DC, and SC layers should have a main absorption band of half an octave in width (factor of 1.5). In this way, two groups of UC/SC/DC ensembles can be accommodated on top of each other, so that the stack consists of six bandgaps in total, see system MEG-2 shown in Fig 2d. As shown in Figure 2a, black line, such a hypothetical two-junction MEG device would outperform a standard 5-junction device at AM1.5, and would be nearly insensitive to spectral changes, still providing over 55% PCE at AM5.0.

### 3b     Real systems: Combining MEG and lateral architectures

In lateral architectures, we want to exploit light management concepts such as phase masks.[4] Relying on diffractive optics, the maximum diffraction angles are given by the sun's solid angle and can exceed 10°, depending on pixel size and wavelength.[4] Commercially competitive PV systems will restrict the distance between phase mask and active layers to 1 mm (in a free-standing device) or 1 cm (façade integrated into a double window), so that the lateral devices, along the axis of light management, will need to have sub-mm dimensions. Such lateral resolution can be achieved with solution-processed PV materials and digital printing[19]. In order to predict maximum achievable PCE values in lateral systems comprising MEG materials, we thus assume the SC to have voltage losses of $\Delta V_{OC} = 0.47\ V$.[20] Under this constraint, we optimize the optical bandgaps of the SC, UC and DC layers as well as the bandpass filters (see ESI, Figure S2) for the incoming solar irradiation to simulate light management by phase masks. As in large-scale devices, the phase masks will be periodic, we take into account an overlap of the tails of the spectral distributions such that the device receiving the shortest wavelengths also will receive part of the long-wavelength illumination. In the following, we show how this fact can be exploited by MEG materials.

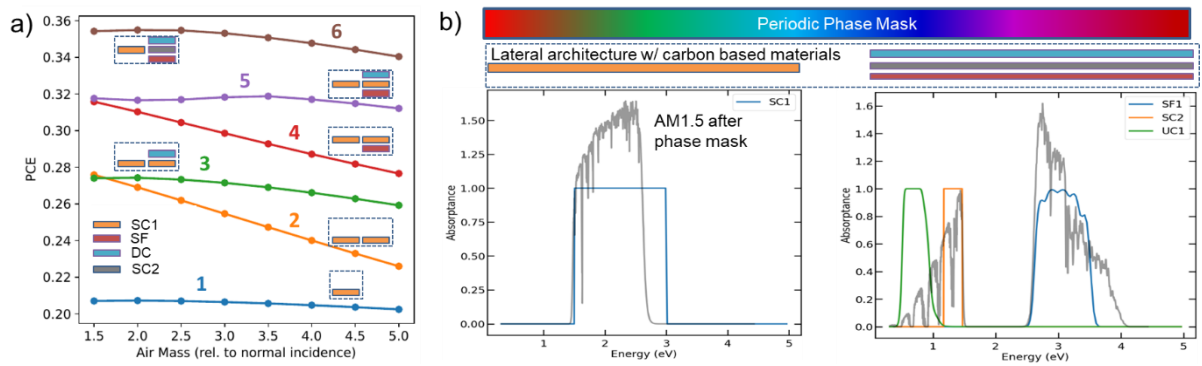

*Figure 3. Simulated PCE of lateral double junction architectures assuming typical parameters for solution processable OPV systems. The performance of a single junction is given for comparison (blue line). Simulation parameters: IQE = 0.9; FF=0.8; $\Delta V_{OC} = 0.47\ V$ except for SC2 which has $\Delta V_{OC} = 0.27\ V$. b). Spectral distribution of a periodic phase mask distributing AM1.5 illumination on the left and right pixel of the lateral double junction. The assumed absorptance spectra of the respective layers are given as colored curves according the legend, the illumination spectra are plotted in grey.*

Figure 3a shows that under the assumptions summarized in the caption of Figure 3, the maximum achievable PCE for a single junction device is close to 21% (blue line), slightly exceeding the actually achieved performance of today's best OPV systems due to the assumption of boxcar absorptance. In order to achieve near-boxcar absorptance, film thicknesses of OPV active layers would need to be increased which can only be done if charge carrier mobilities are improved, because otherwise FF losses would result. Nonetheless, comparison of Fig. 2a ($\Delta V_{OC} = 0.2\ V$) and Fig.3a ($\Delta V_{OC} = 0.47\ V$) suggests that the biggest gain will come from reducing voltage losses, while improving mobilities will yield only a marginal PCE increase.

The orange curve in Figure 3a refers to a lateral double junction, reaching maximum PCE values close to 28% under the assumptions given in the caption of Figure 3. Due to Kirchhoff's rules, the series connection limits $J_{sc}$ of the lateral stack to the $J_{sc}$ produced in the short wavelength device, resulting in severe PCE losses when going from AM1.5 to AM5.0. The resilience to spectral changes is strongly improved if a DC layer is added, see green line in Figure 3a. In this architecture, we exploit a spectral overlap between long and short wavelength regions caused by the periodic phase mask, such that the DC-containing device is receiving both the long and short wavelength portions of the sunlight, while the optimal bandgap of the second device is located at intermediate wavelengths. This is shown by the gray lines in Figure 3b and c. In consequence, reducing blue light (going from AM1.5 to AM5.0) will result in approximately balanced losses in both lateral devices, explaining the improved resilience against spectral changes shown in Fig. 3a. It is important to note that the optimal DC bandgap for this architecture (2.54 eV, see Table S3, ESI) is close to the published tetracene:PbS system[12] such that only slight modifications in chemical structure or layer deposition parameters will suffice for the experimental realisation of this architecture.

Using an UC instead of a DC layer yields significantly higher PCE values, however at the price of a reduced resilience against spectral changes (red line in Fig. 3a). This architecture requires UC systems with 90% upconversion yield, which have so far not been presented under unconcentrated sunlight. The challenge can be approached from two sides: on one hand, excited state lifetimes of UC systems should be increased by avoid loss channels coming from disorder or impurities;[3] on the other hand, phase masks, by means of genetic algorithms, may be optimized to periodically concentrate light absorbed by UC systems as much as possible.

In a lateral architecture containing both UC and DC layers (purple line in Fig. 3a), we obtain near-perfect resilience against spectral changes at PCE values around 32%. This shows that even assuming $\Delta V_{OC} = 0.47\ V$, this architecture can bring PCE of lateral tandems close to the single junction limit for inorganic systems, obtained for $\Delta V_{OC} = 0.2\ V$. Furthermore, this architecture provides a handle to deal with the Energy Gap Law: as the device containing the MEG layers receives mainly short and long but not intermediate wavelengths, the role of the SC material is merely to harvest the narrowband emission of the UC and DC photons. This relaxes the requirement of broadband absorption, usually mandatory in solar materials. We can thus revert to solution printable materials with less electron-phonon coupling, which in standard Bulk heterojunction architectures is needed for broadband absorption. In the extreme case of a hypothetical highly luminescent narrowband absorber, overall voltage losses may approach those given by radiative losses, typically close to 0.25 eV.[21] Here, we assume $\Delta V_{OC} = 0.27\ V$ for the MEG-matched SC2, which amounts to the limiting case of near-unity luminescent quantum efficiency. Assuming $\Delta V_{OC} = 0.47\ V$ for the isolated and 0.27 V for the MEG-matched SC, we predict PCE values close to 36% and near perfect resilience against spectral changes, see brown curve in Figure 3a, see Table S3, ESI.

### 3c    Real systems: virtual layer optimizations using fast GPR surrogates

In order to predict the maximum achievable PCE in MEG architectures comprising realistic systems, we used machine-learned predictions of $V_{OC}$, $J_{SC}$, and FF based on spectral features of the active layer. We chose the system PM6:BTP-4F-12:[70]PCBM because of the availability of a highly reliable experimental dataset from autonomous high throughput optimization.[22] Out of five predictors, we chose to optimize the total absorption and the D:A absorption ratio because these parameters can be directly experimentally controlled by varying the active layer thickness and the donor:acceptor ratio. The table in Figure 4c reports the optimized fit parameters and resulting PCE increase

Next, we assumed a DC material to be radiatively coupled to the SC. As reference DC material, we chose the system tetracene:PbS described in reference [12] because it can be radiatively coupled to the SC which allows theoretically unlimited DC layer thickness. However, the emission of the tetracene:PbS system occurs at 950 nm where the semiconductor does not absorb. Therefore, we allowed the optimizer to tune both the tetracene absorption maximum ($E_{DC}$) and the PbS emission maximum ($E_{PL}$) such that $E_{PL} = \frac{E_{DC}}{2} - 0.24\ eV$.[12] This procedure is equivalent to searching for a hypothetical SF system with optimum energy matching, while retaining the experimentally observed spectral properties and efficiencies of the original tetracene/PbS system.

Figure S4a shows the surrogate function from the Bayesian Optimization. The table in Figure 4c shows that the optimized system DC:SC achieves PCE=15.8%, which is 1% more than the pure SC. The table also shows that in the presence of the DC layer, the optimized D:A blend requires a slightly higher acceptor molar fraction ($X_A$). The reason for this is a cross-correlation between $X_A$ and $E_{DC}$, shown as diagonally oriented orange ellipse (connecting points of same predicted PCE) in Figure S4a. This means that in order to go for lower energies (where the DC material can harvest more photons), $X_A$ must increase. This example shows the capacity of the algorithm to predict an optimized active layer composition based on machine-learned structure property relationships.

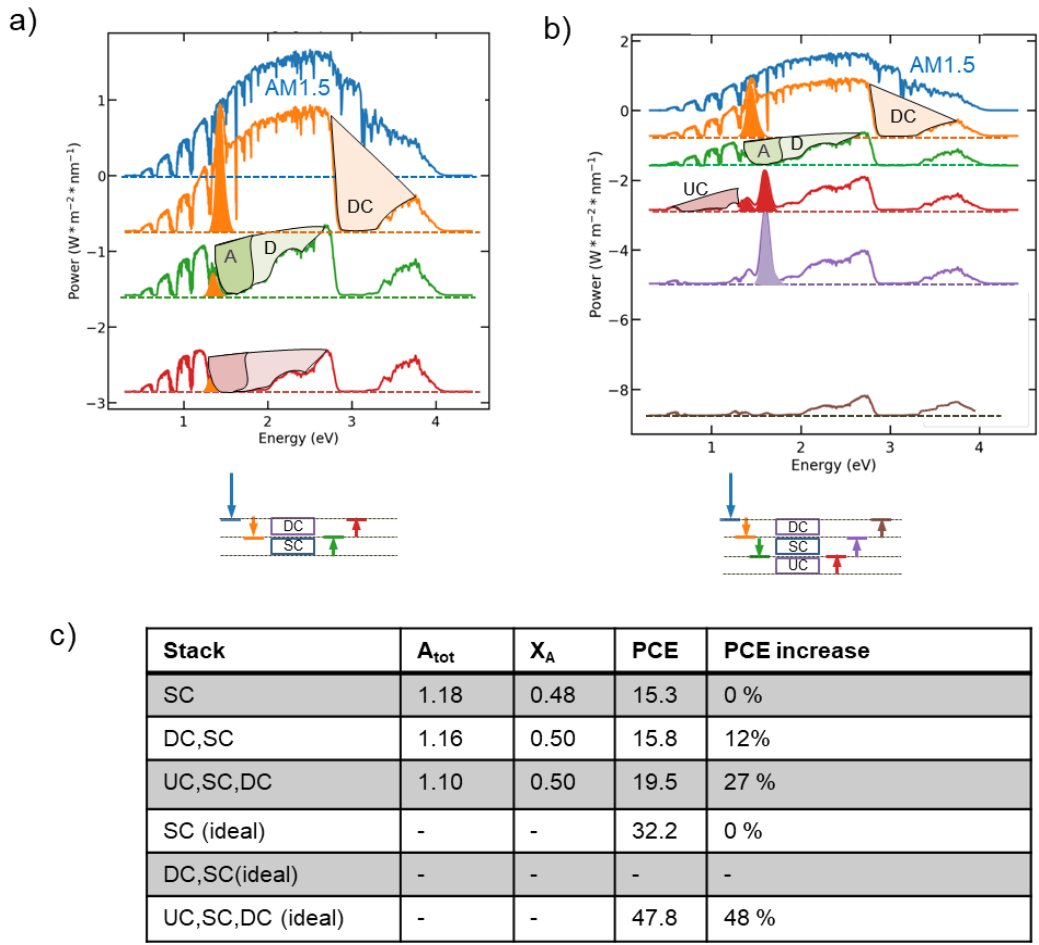

Figure 4. Bayesian Optimization of MEG system using structure-property relationships of PM6:BTP-4F-12:[70]PCBM blends obtained from experimental high throughput by machine learning. a) Spectral distribution of incoming light for each layer for the optimized DC:SC system, in forward direction and after reflection by the bottom electrode (light incident from the top). Faintly shaded areas: absorption by the previous layer; boldly shaded areas: emission by the previous layer. b) same for the system DC:SC:UC. c) optimization parameters and predictions of the electrical performance parameters. The term 'ideal' refers to the ideal systems described in Figure 2. In panels a and b,, "D" refers to PM6 (the donor), while "A" refers to BTP-4F-12 (the acceptor). c) summary table showing the optimized layer composition parameters $A_{tot}$ and $X_A$, which stand for the total thickness and the donor:acceptor ratio, together with the predicted PCE and its relative increase due to the inclusion of MEG layers.

Figure 4a shows the spectral distribution of the incident light for each layer. The optimum energy of the down conversion layer is predicted at 2.91 eV, which is more than half an eV blue shifted against the original tetracene absorption. It might be challenging to achieve such a strong blueshift by modifying the tetracene molecule while still retaining the condition that triplet energies be half of the singlet energies. An alternative approach may be to choose a more red-shifted OPV system; however even non-fullerene acceptors like Y11 with an optical bandgap of 1.31 eV would not match the tetracene system.[23] Figure 4a shows that due to the high DC film thickness, there is complete absorption of the incident light from 2.9-3.2 eV. However, the spectral region from 3.3-4.0 eV is not absorbed and would proceed into the SC layer, causing thermalization losses (and possibly UV degradation).[24] It is interesting to note that the emission of the PbS (peak at 1.4 eV in orange spectrum in Figure 4b) is not in the maximum, but in the edge of SC absorption, such that it is fully absorbed only after the second pass through the SC. This shows the ability of the Bayesian Optimizer to find a condition allowing maximum light harvesting (lower optical bandgap $E_{opt}^0$) while respecting the film thickness limits for efficient charge extraction in the SC layer.

Next, we add a generic UC material as bottom layer. In Figure 4b, we show the corresponding spectral distribution. Assuming 100% UC yield, the PCE of the combined DC:SC:UC device would be 18.7%. This is not even half of the PCE values predicted for the ideal, boxcar system, see gray curve in Figure 2a. In order to distinguish losses in the SC from those caused by the MEG layers, we look at the relative increase brought by the MEG materials. For the realistic systems, this relative increase is 27% while in Figure 3a, the relative increase is 48%. There are several reasons that contribute to the lower performance of the realistic systems. One is incomplete absorption causing $J_{sc}$ losses: the bottom curves in both figures 4c and d still show a substantial amount of non-absorbed photons that will be reflected . A further reason for $J_{sc}$ losses is the non-unity PbS emission efficiency (90%,[12]), of which only around 80% are forward transmitted into the SC layer.[25] Moreover, there are energy losses in the PbS system which emits not at exactly half the tetracene absorption but 0.24 eV below; in order to match the SC bandgap (which defines the achievable $V_{OC}$), $E_{DC}$ must be higher by 0.24 eV than in an ideal boxcar system, where the solar spectrum has less intensity, hence another portion of $J_{sc}$ is lost.

Still, a relative increase of PCE by 24% induced by realistically performing MEG materials seems promising. The main limitation on the SC side is the high non-radiative voltage losses of the machine learned PM6:BTP-4F-12 system, amounting to 0.65V. Furthermore, no UC system has so far been described showing 100% UC yield under 1 sun.

Summarizing, our simulations using realistic systems show that UC and DC processes have a realistic potential to contribute to a performance increase of emerging PV materials beyond the detailed balance limit. It matches the possible performance increase in tandem systems, exceeding the performance of the latter in terms of resilience against spectral changes of the solar irradiation. Given the superior ease of processing of MEG systems compared to tandem cells, these results encourage research to resolve the existing challenges in MEG systems. Moreover, the lateral architectures simulated in Figure 3 can in principle be processed by upscalable techniques such as digital printing.

## 4    Conclusions

We have performed Bayesian Optimization to predict and optimize the electrical performance of multi-junction architectures, both vertical and lateral, in combination with multi-exciton materials, by using prior knowledge about the deployed materials. This knowledge can be either in the form of published transfer yields and spectral shapes but also in the form of quantitative structure-property relationships (QSPR) obtained by machine learning. We have found that by optimizing bandgap energies of multi-exciton generation (MEG) layers, double junction vertical stacks can reach efficiencies beyond those of five-junction tandem devices. Moreover, such combinations of MEG and double junction devices would be highly resilient against spectral changes of the incoming sunlight. We have simulated such double junctions using combinations of real OPV systems from our database, and quantified the most promising pathways to further improve PCE. We have further found that MEG layers in lateral architectures can improve resilience against spectral changes and might allow reducing nonradiative voltage losses following the Energy Gap Law. Finally, we have shown that the simulation environment is able to use machine.-learned QSPR from high throughput experiments to virtually optimise the active layer (such as, the film thickness and the donor-acceptor ratio) for a given architecture. The simulation environment thus represents an important building block towards a digital twin of PV materials.

## 5 Acknowledgments

Financial support from the DFG is gratefully acknowledged (BR 4031/22-1 and BR 4031/21-1).

**References:**

[1] https://www.ise.fraunhofer.de/content/dam/ise/de/documents/publications/studies/Photovoltaics-Report.pdf

[2] https://www.nrel.gov/pv/interactive-cell-efficiency.html

[3] Bernhard Mitchell, Gerhard Peharz, Gerald Siefer, Marius Peters, Tobias Gandy, Jan Christoph Goldschmidt, Jan Benick, Stefan W. Glunz, Andreas W. Bett, Frank Dimroth, Four-junction spectral beam-splitting photovoltaic receiver with high optical efficiency, Prog. Photovolt: Res. Appl.2011;19:61–72

[4] T. Patrick Xiao, Osman S. Cifci, Samarth Bhargava, Hao Chen, Timo Gissibl, Weijun Zhou, Harald Giessen, Kimani C. Toussaint, Jr., Eli Yablonovitch, Paul V. Braun, Diffractive Spectral-Splitting Optical Element Designed by Adjoint-Based Electromagnetic Optimization and Fabricated by Femtosecond 3D Direct Laser Writing, ACS Photonics 2016, 3, 886–894, DOI: 10.1021/acsphotonics.6b00066

[5] Qingli Huang, Qi Peng, Jianyao Hu, Huawei Xu, Chunxu Jiang, Qunxing Liu, Design of a high-efficiency and low-cost reflection-type diffractive optical element as the spectrum splitting solar concentrator for lateral multi-junction solar cells architecture, 2016 IEEE Advanced Information Management, Communicates, Electronic and Automation Control Conference (IMCEC), Xi'an, China, 2016, pp. 1528-1532, doi: 10.1109/IMCEC.2016.7867473.

[6] artí Gibert-Roca, Miquel Casademont-Viñas, Quan Liu, Koen Vandewal, Alejandro R. Goñi, and Mariano Campoy-Quiles, RAINBOW Organic Solar Cells: Implementing SpectralSplitting in Lateral Multi-Junction Architectures, Adv. Mater.2023, 2212226, DOI: 10.1002/adma.202212226

[7] Tim F. Schulze and Timothy W. Schmidt, Photochemical upconversion: present status and prospects for its application to solar energy conversion, Energy Environ. Sci., 2015, 8, 103

[8] Nienhaus, L., Wu, M., Geva, N., Shepherd, J.J., Wilson, M.W.B., Bulović, V., Van Voorhis, T., Baldo, M.A., and Bawendi, M.G.: Speed limit for triplet-exciton transfer in solid-state pbs nanocrystal-sensitized photon upconversion. ACS Nano11, 7848(2017)

[9] Tobias Ullrich, Dominik Munz and Dirk M. Guldi, Unconventional singlet fission materials, Chem. Soc. Rev., 2021, 50, 3485

[10] Jianlong Xia, Samuel N. Sanders, Wei Cheng, Jonathan Z. Low, Jinping Liu, Luis M. Campos, and Taolei Sun, Singlet Fission: Progress and Prospects in Solar Cells, Adv. Mater. 2017, 29, 1601652

[11] Park, S., Roy, A., Beaupré, S. et al. Bulk heterojunction solar cells with internal quantum efficiency approaching 100%. Nature Photon 3, 297–302 (2009). https://doi.org/10.1038/nphoton.2009.69

[12] Nicholas J. Thompson, Mark W. B. Wilson, Daniel N. Congreve, Patrick R. Brown, Jennifer M. Scherer, Thomas S. Bischof, Mengfei Wu, Nadav Geva, Matthew Welborn, Troy Van Voorhis, Vladimir Bulović, Moungi G. Bawendi and Marc A. Baldo, Energy harvesting of non-emissive triplet excitons in tetracene by emissive PbS nanocrystals, Nature Mater 13, 1039–1043 (2014). https://doi.org/10.1038/nmat4097

[13] Eshwar Ravishankar, Ronald E. Booth, Joseph A. Hollingsworth, Harald Ade, Heike Sederoff, Joseph F. DeCarolis and Brendan T. O'Connor, Energy Environ. Sci., 2022,15, 1659-1671

[14] Traverse, C.J., Pandey, R., Barr, M.C. et al. Emergence of highly transparent photovoltaics for distributed applications. Nat Energy 2, 849–860 (2017). https://doi.org/10.1038/s41560-017-0016-9

[15] https://github.com/i-MEET/boar

[16] a) S. T. Turner, P. Pingel, R. Steyrleuthner, E. J. W. Crossland, S. Ludwigs, D. Neher, Adv. Funct. Mater. 2011, 21, 4640; b) F. C. Spano, J. Chem. Phys. 2005, 122, 234701; c) J. Clark, C. Silva, R. H. Friend, F. C. Spano, Phys. Rev. Lett. 2007, 98, 206406

[17] Osbel Almora, Derya Baran, Guillermo C. Bazan, Carlos I. Cabrera, Sule Erten-Ela, Karen Forberich, Fei Guo, Jens Hauch, Anita W. Y. Ho-Baillie, T. Jesper Jacobsson, Rene A. J. Janssen, Thomas Kirchartz, Nikos Kopidakis, Maria A. Loi, Richard R. Lunt, Xavier Mathew, Michael D. McGehee, Jie Min, David B. Mitzi, Mohammad K. Nazeeruddin, Jenny Nelson, Ana F. Nogueira, Ulrich W. Paetzold, Barry P. Rand, Uwe Rau, Henry J. Snaith, Eva Unger, Lídice Vaillant-Roca, Chenchen Yang, Hin-Lap Yip, Christoph J. Brabec, Device Performance of Emerging Photovoltaic Materials (Version 3), Adv. Energy Mater. 2023, 13, 2203313

[18] M. C. Hanna and A. J. Nozik, Solar conversion efficiency of photovoltaic and photoelectrolysis cells with carrier multiplication absorbers, J. Appl. Phys. 100, 074510, 2006

[19] M. -K. Hamjah, M. Steinberger, K. C. Tam, H. -J. Egelhaaf, C. J. Brabec and J. Franke, "Aerosol jet printed AgNW electrode and PEDOT:PSS layers for organic light-emitting diode devices fabrication," 2021 14th

International Congress Molded Interconnect Devices (MID), Amberg, Germany, 2021, pp. 1-4, doi: 10.1109/MID50463.2021.9361616

[20] Wu, Q., Wang, W., Wang, T. et al. High-performance all-polymer solar cells with only 0.47 eV energy loss. Sci. China Chem. 63, 1449–1460 (2020). https://doi.org/10.1007/s11426-020-9785-7

[21] Johannes Benduhn, Kristofer Tvingstedt, Fortunato Piersimoni, Sascha Ullbrich, Yeli Fan, Manuel Tropiano, Kathryn A. McGarry, Olaf Zeika, Moritz K. Riede, Christopher J. Douglas, Stephen Barlow, Seth R. Marder, Dieter Neher, Donato Spoltore and Koen Vandewal, Intrinsic non-radiative voltage losses in fullerene-based organic solar cells, Nat. Energy 2, 17053 (2017), DOI: 10.1038/nenergy.2017.5

[22] Osterrieder, T., Schmitt, F., Lüer, L., Wagner, J., Heumüller, T., Hauch, J., Brabec, C.J., "Autonomous Optimization of an Organic Solar Cell in a 4-dimensional Parameter Space", arXiv:2305.08248 (2023)

[23] Wei Gao, Francis R. Lin, Alex K.-Y. Jen, Near-Infrared Absorbing Nonfullerene Acceptors for Organic Solar Cells, Sol. RRL 2022, 6, 2100

[24] Paul Weitz, Vincent Marc Le Corre, Xiaoyan Du, Karen Forberich, Carsten Deibel, Christoph J. Brabec, and Thomas Heumüller, Revealing Photodegradation Pathways of Organic Solar Cells by Spectrally Resolved Accelerated Lifetime Analysis, Adv. Energy Mater. 2023, 13, 2202564

[25] Taylor Uekert, Anastasiia Solodovnyk, Sergei Ponomarenko, Andres Osvet, Ievgen Levchuk, Jessica Gast, Miroslaw Batentschuk, Karen Forberich, Edda Stern, Hans-Joachim Egelhaaf, Christoph J. Brabec, Nanostructured organosilicon luminophores in highly efficient luminescent down-shifting layers for thin film photovoltaics, Sol. En. Mat. Sol. Cells 155 (2016) 1-8

# Bypassing the single junction limit with advanced photovoltaic architectures

Larry Lüer[1], Marius Peters[2], Dan Bornstein[1], Vincent M. Le Corre[1], Karen Forberich[2], Dirk Guldi[3], Christoph Brabec[1,2]

[1] Institute of Materials for Electronics and Energy Technology (i-MEET), Friedrich-Alexander-Universität Erlangen-Nürnberg, Martensstrasse 7, 91058 Erlangen, Germany

[2] High Throughput Methods in Photovoltaics, Forschungszentrum Jülich GmbH, Helmholtz Institute Erlangen-Nürnberg for Renewable Energy (HI ERN), Immerwahrstraße 2, 91058 Erlangen, Germany

[3] Department of Chemistry and Pharmacy, Egerlandstr. 3, 91058 Erlangen, Germany

**ELECTRONIC SUPPORTING INFORMATION (ESI)**

**Table of contents:**

## A    THE SIMULATION AND OPTIMIZATION ENVIRONMENT

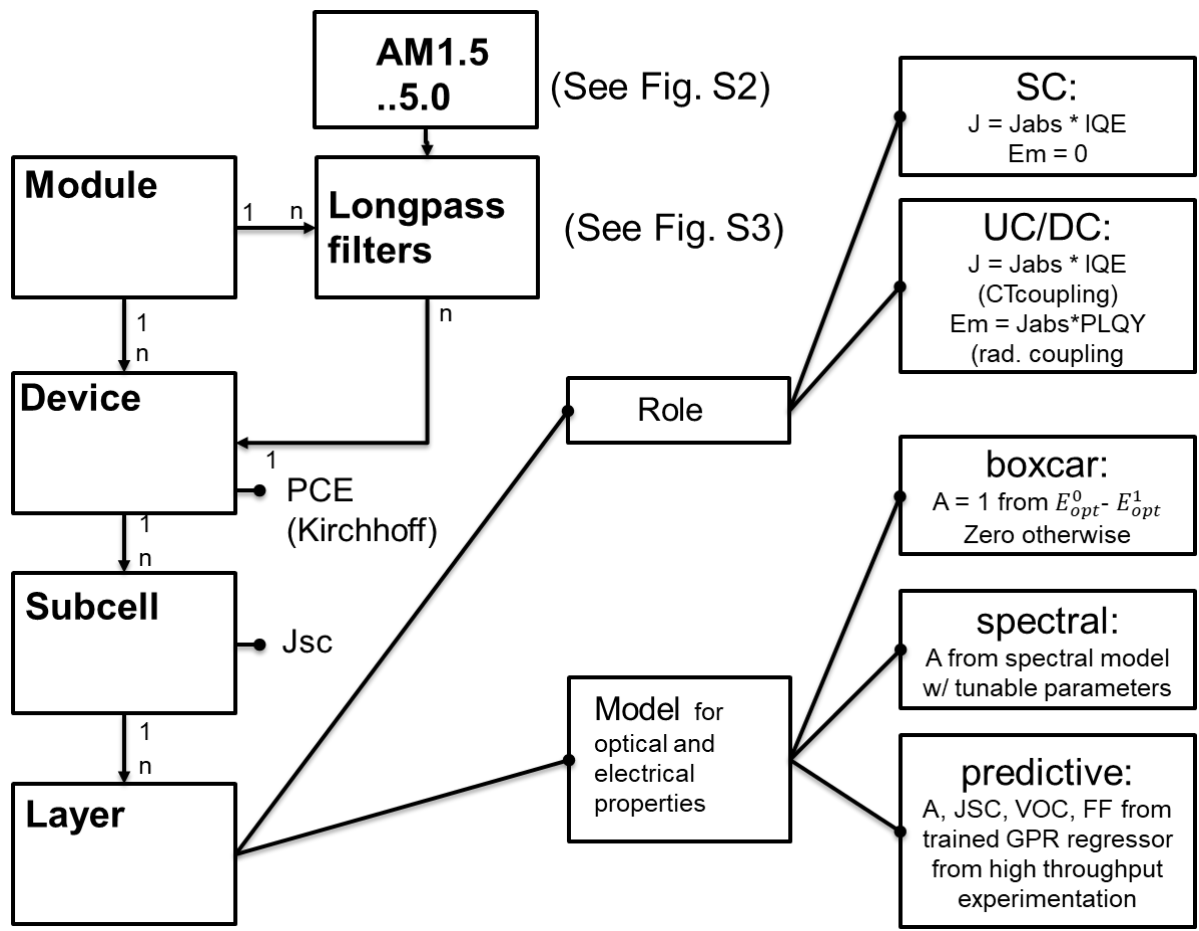

Figure S1. Structure of the PV system in the simulation environment for both vertical and lateral stacks subject to spectral sharing schemes. A = Absorptance; arrows 1:n mean: one origin can relate to n targets. Em=emission

In Figure 2, left side, we show the Python classes that represent the structure of the PV system. The basic entity is a layer, which can be of role "SC", "UC", or "DC". Each layer is associated with a spectral model, which can be "boxcar", "spectral" or "predictive". Several UC or DC layers, but only and exactly one SC layer, make up a subcell. On the subcell level, the electrical parameters are obtained. If the spectral model is "boxcar" or "spectral", then the photocurrent is obtained by multiplying the incoming spectral irradiance with the absorptance of the SC layer, and assuming fixed values for the internal quantum efficiency (IQE) and the fill factor (FF), to match available experimental data. The open circuit voltage ($V_{OC}$) is obtained by subtracting a quantity $\Delta V_{OC}$ from the optical bandgap to match available experimental data. If the model is "predictive", then we use machine learned predictions of $V_{oc}$, FF, and $J_{SC}$, as function of process conditions, in the form of trained Gaussian Process Regressors from experimental High Throughput studies of OPV devices.[26] As these regressors have been trained under one sun illumination and do not contain trends with respect to light intensity, we simply scale the predicted $J_{sc}$ values with the incoming intensity. We highlight that interference effects, usually described with the transfer matrix method (TMM) are implicitly considered in the trained regressors, because the film thicknesses have been varied over a wide range during the HT experiments. The contribution of the DC and UC layers to the $J_{sc}$ of a subcell, in charge coupling mode, occurs via multiplying the excitation current in the DC or UC layer with a charge transfer efficiency, which is generally unknown and can be freely chosen up to unity efficiency. In radiative transfer mode, emission is triggered multiplying the excitation current with the emission quantum yield (PLQY) for which literature data exist. In internal layers without contact to air, half of the emission spectrum is added to the incoming light of the next layer, while the other half is considered when the excitation current from the reflected light is calculated, that is, when the incoming light is travelling in the opposite direction. In case the emissive layer is at the top, then 81% of the emission spectrum is projected forward, and 19% are emitted into the vacuum.[27] We furthermore use the SMARTS 2.9.5 software[28] to calculate the global tilted irradiance solar spectrum at air mass (AM) 1.5 – 5.0, using the same atmospheric conditions as the ASTM G173-03 reference spectra.

One or more subcells can be organized into devices, which hold the vertical architecture of the device. Kirchhoff rules are approximately enforced by summing up the $V_{oc}$ values of the single subcells, and using the minimum $J_{sc}$ and FF of all subcells as values for the device.

Finally, devices can be organized into one and exactly one module, which represents the lateral architecture of the PV system. Only linear arrangements of devices are considered so far. Kirchhoff rules are enforced in the same way as in devices. The most important aspect on the module level is light management. We assume a dispersion of the solar spectrum (see Figure S2) across the devices by defining "dividers" (long pass filters) which are assumed as sigmoid functions with center wavelength and width (of generally 10 nm) accounting for incomplete spectral separation. The resulting spectral bands can be made impinging on any of the devices, as shown schematically in Figure S3. One spectral band can impinge on exactly one device, but one device can receive spectral bands from more than one divider. A periodic spectral modulation can thus be simulated by simply connecting the last, most infrared spectral band, to the first device, receiving the shortest wavelengths.

For any given architecture, we optimize the resulting PCE by varying the free parameters (bandgap energies, spectral parameters, layer thicknesses and donor:acceptor ratios) using BOAR,[29] the Bayesian Optimizer for Automated Research, which is a Python library on top of the scikit-optimize Python package[30] able to optimize up to 10 free parameters and yield parameter interdependences by sampling the posterior distribution. As surrogate function we use a Matern 3/2 kernel being able to match abrupt changes of PCE at band edges.

B    Solar irradiation

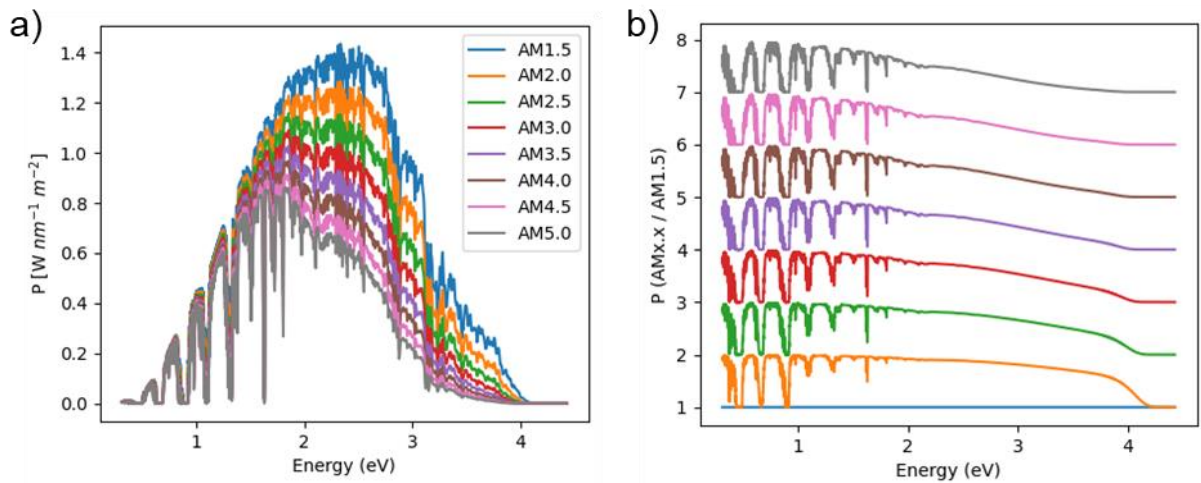

Figure S2: Spectral density of solar irradiation according to SMARTS 2.9.5

We use the SMARTS 2.9.5 software[2] to calculate the global tilted irradiance solar spectrum at air mass (AM) 1.5 – 5.0, using the same atmospheric conditions as the ASTM G173-03 reference spectra. Figure S2a shows the calculated spectral densities used in the manuscript. Note that the spectra are given in units of W $nm^{-1}$ $m^{-2}$, while the X axis is on an energy scale. Greater air mass reduces the spectral density at higher photon energies due to scattering, but also increases absorption by the water bands in the near infrared spectral region, as can be seen more clearly in Figure S2b, where individual spectral densities are shown relative to AM1.5 and shifted vertically for clarity of presentation. Same color scale applies as Figure S2a.

C    Spectral management

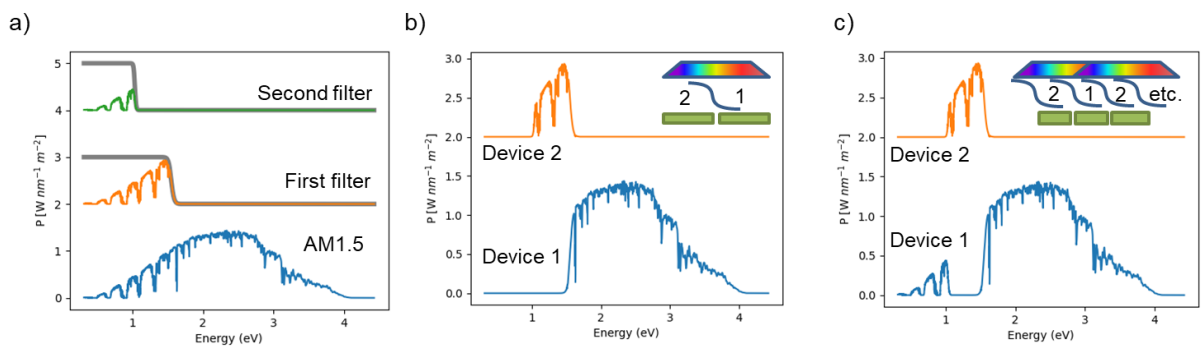

Figure S3: Simulation of light management for lateral devices A) subsequent application of longpass filters (at 1200 and 800 nm, respectively), simulated as sigmoidal functions, b) resulting spectral density of incoming illumination for a two-junction lateral module (depicted in the inset) and for a periodic modulation.

In non-periodic spectral management (panel b), remaining spectrum after second filter (green curve in panel a) is discarded, in periodic spectral management (panel c), it is added to device 1. "Personalized spectral management" is also possible by the 1:n connection shown in Figure S1, allowing complete freedom in spectral design for each device. This feature can be used to model light management beyond phase masks, e.g. using nanophotonics approaches, able to achieve higher light concentration factors than phase masks.

**D     Parameters for optimized systems**

| Number ands stack | PCE, AM1.5 (%) | PCE, AM5.0 (%) | $E_g$ (eV) |
|---|---|---|---|
| 1 (1 x SC) | 32.2 | 33.3 | 1.135 |
| 2 (2 x SC) | 43.8 | 38.8 | 1.59, 0.94 |
| 3 (3 x SC) | 49.1 | 39.8 | 1.79, 1.19, 0.68 |
| 4 (4 x SC) | 53.1 | 39.3 | 1.98, 1.47, 1.06, 0.70 |
| 5 (5 x SC) | 54.8 | 38.5 | 2.1, 1.61, 1.24, 0.94, 0.56 |
| MEG 1: (SC,UC,DC} | 47.8 | 48.3 | 1.38, 2.76, 0.69 |
| MEG2: SC,UC,DC \| SC,UC,DC | 56.3 | 55.1 | 0.695, 0.925, 1.39, 1.85, 2.78, 3.7 |

Table S1: PCE values at AM1.5 and AM5.0, and optimized bandgap energies for the vertical architectures in Figure 2a. Further parameters were: IQE = 1; FF=0.8; $\Delta V_{OC} = 0.2V$.

| Number and stack | PCE, AM1.5 (%) | PCE, AM5.0 (%) | $E_g$ (eV) |
|---|---|---|---|
| 1 (1 x SC) | 32.0 | 33.0 | 1.13 |
| 2 (2 x SC) | 43.2 | 38.1 | 1.59, 0.93 |
| 3 (3 x SC) | 48.5 | 39.3 | 1.78, 1.19, 0.69 |
| 4 (4 x SC) | 50.8 | 37.0 | 1.99, 1.47, 1.06, 0.67 |
| 5 (5 x SC) | 52.0 | 36.2 | 2.1, 1.60, 1.22, 0.89, 0.53 |

Table S2: PCE values at AM1.5 and AM5.0, and bandgap energies for the lateral architectures in Figure 2b. Further parameters were: IQE = 1.0; FF=0.8; $\Delta V_{OC} = 0.2 V$.

| Number and stack | PCE, AM1.5 (%) | PCE, AM5.0 (%) | $E_g$ (eV) |
|---|---|---|---|
| 1: SC | 20.7 | 20.2 | 1.38 |
| 2: SC|SC | 27.6 | 22.6 | 1.77, 1.16 |
| 3: SC|(SC, DC) | 27.4 | 25.9 | 1.58, 1.14, 2.54 |
| 5: SC |(SC, DC, UC) | 31.8 | 31.2 | 1.49, 1.297, 2.78, 0.70 |
| 4: SC | (SC, UC) | 31.6 | 27.7 | 1.20, 1.83, 0.96 |
| 6: SC | (SC, UC, DC) | 35.4 | 34.0 | 1.53, 1.29, 2.80, 0.70 |

Table S3: PCE values at AM1.5 and AM5.0, and bandgap energies for the simulations in Figure 3. Further parameters were: IQE = 0.9; FF=0.8; $\Delta V_{OC} = 0.47V$ except in last row where second SC has $\Delta V_{OC} = 0.27V$

## E Objective Functions

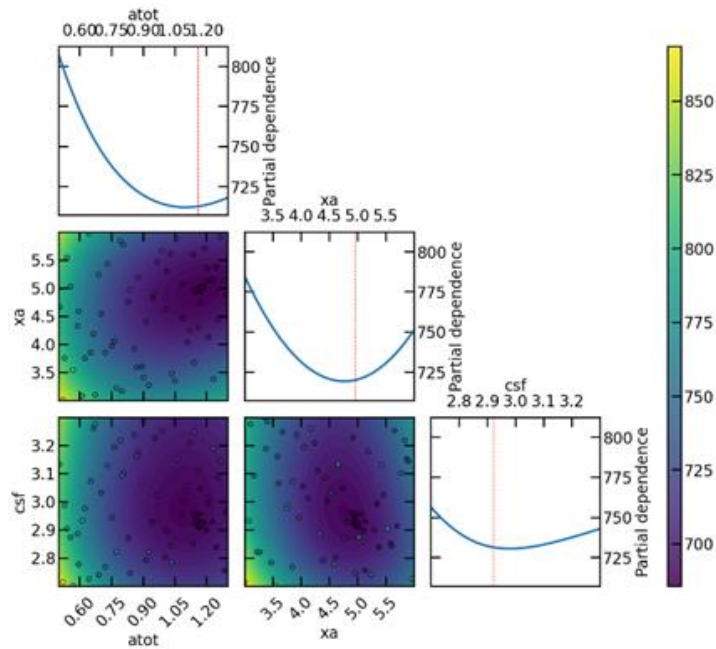

**Figure S4:** Surrogate function obtained from the Bayesian optimization of SC thickness, D:A ratio and DC absorption band gap, given as false color hypersurface. Orange ellipses indicate cross correlation between the optimization parameters. One-dimensional partial dependence plots are given on the diagonal (arbitrary units). Since the Bayesian Optimization is implemented to minimize a certain value, the difference to an arbitrary unrealistically high PCE value (70%) was chosen as the optimization target.

[26] Osterrieder, T., Schmitt, F., Lüer, L., Wagner, J., Heumüller, T., Hauch, J., Brabec, C.J., Autonomous Optimization of an Organic Solar Cell in a 4-dimensional Parameter Space, arXiv:2305.08248 (2023)

[27] Taylor Uekert, Anastasiia Solodovnyk, Sergei Ponomarenko, Andres Osvet, Ievgen Levchuk, Jessica Gast, Miroslaw Batentschuk, Karen Forberich, Edda Stern, Hans-Joachim Egelhaaf, Christoph J. Brabec, Nanostructured organosilicon luminophores in highly efficient luminescent down-shifting layers for thin film photovoltaics, Solar Energy Materials and Solar Cells, 155, (2016), 1-8

[28] https://www.nrel.gov/grid/solar-resource/smarts.html

[29] https://github.com/i-MEET/boar/

[30] https://scikit-optimize.github.io/stable/index.html